\documentclass[prl,twocolumn,showkeys]{revtex4-1}
  \usepackage{graphicx}
  \usepackage{amsmath}
  \usepackage{amssymb}
  \usepackage{makeidx}
  \usepackage{amsfonts}
  \usepackage{appendix}
  \usepackage{mathrsfs}
  \usepackage[usenames,dvipsnames]{pstricks}
	\usepackage{epsfig}
	\usepackage{pst-grad} 
	\usepackage{pst-plot} 
	\usepackage[colorlinks,hyperindex]{hyperref}  
  \hypersetup
  {
    colorlinks,%
    citecolor=black,%
    linkcolor=black,%
    urlcolor=black,%
  }

\makeindex

\begin{document}

\title { Dynamics of Interacting Hotspots-II}

\author{Suman Dutta}
\email{sumand@imsc.res.in}

\affiliation{The Institute of Mathematical Sciences\\
CIT Campus, Taramani, Chennai 600113 \\ India.}

\date{20 May, 2015 }

\begin{abstract}

In the absence of any proper clinical solution, human civilization is only left with sophisticated intervention measures to contain the spread of COVID-19. However, the existing models to estimate the intervention does not take into account the realistic connectivity of the epicentres of the pandemic. We generalise our earlier model of {\it interacting hotspots} to test various possibilities of intervention in a model state consisting of multiple epicentres. We also analyse situations when the hotspots are spatially correlated and the interaction is limited to population exchanges with the nearest neighbours. We show that the heterogeneity in the infection propagation is solely dependent on the protocol of the containment and its strength.  We explore many such situations and discuss possibilities.\\

\end{abstract}
\date{\today~  }

\maketitle

The transportation of the infected persons can lead to large scale community transmissions during a pandemic situation\cite{lancet1}. Therefore, to avoid world-wide transmission, most of the countries have sealed the borders with their neighbours, stopping every-other non-essential activities for more than a fortnight\cite{science3}. However, millions of poeple have travelled from China, and through different countries, to every part of the globe between the very first report of the disease from Wuhan and the date of imposition of lockdown at a global scale\cite{lancet1,science3}. 

The simplest approach to predict the spread of such pandemic uses familiar methods of inflammatory response model of epidemic that considers deterministic temporal evolution of epidemic variables including different categories of states\cite{rmp}. For instance, the standard SIR model uses three states: A person in susceptible (S) category can pass to a category Infected (I) at a rate $\beta$ while the infected person can pass to a category Recovered (R) at a rate $\gamma$. In recent times, researchers have developed different multi-state models that takes into account a range of complex variables and numerous possibilities of intermediate states\cite{science3,ind1,ger1}. On the other hand, stochastic \cite{pre1,pnas1} and data-driven approaches\cite{science4} use network based analysis to explore the role of stochastic transportation in the dispersal of the virus. Combination of both of the techniques sometimes give surprisingly good agreement with the data \cite{pnas1,science4}.     

Therefore, realistic theoretical models genuinely need to consider the inclusion of realistic human mobility between the regions of exposure in order to estimate the spatial spread of the epidemic at a larger scale. This is an extremely challenging task as every traffic flow between the communities in a region creates local disturbances, jeopardizing the global balance. Maintaining the global conservation away from equilibrium further needs assurance of dynamical equilibrium at a global scale, even when local regions are destabilized by each of the inter-community movements. 

Here, we explore various possibilities of community transmission in presence of intervention, generalising our earlier model of {\it interacting hotspots}\cite{ih1}. We propose a set of generalised deterministic equations with inter-community transport that govern the spread in epidemic. The community transport, eventually, is realised via stochastic exchanges of populations between regions, maintaining a conservation at local scale. The set of stochastic moves successfully ensure long time global equilibrium. We show that complexity of the transport protocol govern the heterogeneity in the infection propagation that defines the overall scale of containment. We also discuss the nature of mitigation process when such hotspots are spatially correlated at nearest neighbour. The model can be easily integrated with most of the epidemic models that exist in the literature.

We start with Susceptible-Infected-Recovered (SIR) model that considers the transition from $S$ to $I$ and $I$ to $R$  with rates $\beta$ and $\gamma$ respectively. Taking into account  the newborns and the natural deaths, the evolution equation for a hotspot becomes 

\begin{equation} \label{GrindEQ__1_} 
\frac{dS}{dt}=-\frac{\beta}{N} SI+\xi N-\lambda S 
\end{equation} 

where $\xi $ is the rate of growth of population due to addition of newborns and $\lambda $ is the rate of death of susceptible persons, and also of persons who have recovered from the epidemic, due to other natural causes.

\begin{equation} \label{GrindEQ__2_} 
\frac{dI}{dt}=\frac{\beta}{N} SI-\gamma I-\zeta I 
\end{equation} 

where $\zeta $ is the rate of death of infected persons. This also includes the natural deaths. $\gamma$ is the recovery rate of the infected patients. 

\begin{equation} \label{GrindEQ__3_} 
\frac{dR}{dt}= \gamma \ I-\lambda R. 
\end{equation} 

Thus,
\begin{equation} \label{GrindEQ__4_} 
\frac{dN}{dt}=\frac{dS}{dt}+\frac{dI}{dt}+\frac{dR}{dt}=\xi N-\lambda (S+R)- \zeta I 
\end{equation} 

Now we think of a model system of {\it interacting hotspots} that exchange population due to inter-city transportation\cite{ds}. Let there be $M\ $number of cities, each identified by an index $i\left(i=1,\ 2,\ 3,\ \dots \dots .,M\right).$ Eqs.\eqref{GrindEQ__1_} -\eqref{GrindEQ__4_} are to be modified as follows:

\begin{equation} \label{GrindEQ__5_} 
\frac{dS^i}{dt}=-\frac{\beta}{N}S^iI^i+\xi N^i-\lambda S^i+\ \sum^M_{j=1}{f^{ij}}S^j-\left(\sum^M_{j=1}{f^{ji}}\right)S^i 
\end{equation} 

where $f^{ij}N^j$ represents the rate of arrival of people from city `$j$' to city `$i$', while $f^{ji}N^i$ is the rate of the reverse process. Obviously, $f^{kl}\ll 1;;k,l=1,\ 2,\ \dots .n$. We note that \noindent $\left({\rm a}\right)f^{ii}=0$; (b) $f^{ij}\ne f^{ji}$, in general; (c) $\beta ,\ \gamma ,\ \xi ,\ \lambda ,\ \zeta $ may depend on index number $i$.

\begin{equation} \label{GrindEQ__6_} 
\frac{dI^i}{dt}=\frac{\beta}{N^{i}}S^iI^i-\gamma I^i-\zeta I^i 
\end{equation} 

\begin{equation} \label{GrindEQ__7_} 
\frac{dR^i}{dt}= \gamma \ I^i-\lambda R^i. 
\end{equation} 

\begin{equation} \label{GrindEQ__8_} 
\sum^M_{i=1}{\frac{dN^i}{dt}}=\sum^M_{i=1}{\left[\xi N^i-\lambda (S^i+R^i)- \zeta I^i\right]} 
\end{equation} 

Note that 
\begin{equation} \label{GrindEQ__9_} 
\sum^M_{i=1}{\left[\sum^M_{j=1}{f^{ij}}S^j-\left(\sum^M_{j=1}{f^{ji}}\right)S^i\right]}=0 
\end{equation} 
Going a few steps further, one may change Eq. \eqref{GrindEQ__6_} to

\begin{equation} \label{GrindEQ__10_} 
\frac{dI^i}{dt}= \frac{\beta}{N^{i}} S^iI^i- \gamma I^i- \zeta I^i+\ \sum^M_{j=1}{g^{ij}}I^j-\left(\sum^M_{j=1}{g^{ji}}\right)I^i 
\end{equation} 

\noindent with 
\begin{equation} \label{GrindEQ__11_} 
\sum^M_{i=1}{\left[\sum^M_{j=1}{g^{ij}}I^j-\left(\sum^M_{j=1}{g^{ji}}\right)I^i\right]}=0 
\end{equation} 

and, similarly, change Eq. \eqref{GrindEQ__7_} to

\begin{equation} \label{GrindEQ__12_} 
\frac{dR^i}{dt}= \gamma \ I^i-\lambda R^i+\ \sum^M_{j=1}{h^{ij}}R^j-\left(\sum^M_{j=1}{h^{ji}}\right)R^i. 
\end{equation} 

and
\begin{equation} \label{GrindEQ__13_} 
\sum^M_{i=1}{\left[\sum^M_{j=1}{h^{ij}}R^j-\left(\sum^M_{j=1}{h^{ji}}\right)R^i\right]}=0 
\end{equation} 

Data show that, globally, from the total number of persons infected by SARS -- COV -2 up to a certain day, 12 -15\% have recovered and 3\% have died. Thus $\zeta <<\gamma $, and we may, for a first approximation ignore the term $\zeta I^i$ in Eqs. \eqref{GrindEQ__6_} and \eqref{GrindEQ__10_}. Data also reveal that in many countries (especially, in Europe) the total population has remained more or less constant over the last decade. Thus we may assume that $\xi N^i-\lambda S^i\approx 0$. We also note that $\eta R^i\ll \ \gamma \ I^i$ ; so the term $\lambda R^i$ may be dropped from \eqref{GrindEQ__7_} and \eqref{GrindEQ__12_}. Our equations then reduce to 

\begin{equation} \label{GrindEQ__15_} 
\frac{dS^i}{dt}=-\frac{\beta}{N^{i}}S^iI^i+\ \sum^M_{j=1}{f^{ij}}S^j-\left(\sum^M_{j=1}{f^{ji}}\right)S^i 
\end{equation} 

\begin{equation} \label{GrindEQ__16_} 
\frac{dI^i}{dt}=\frac{\beta}{N^{i}}S^iI^i-\gamma I^i+\ \sum^M_{j=1}{g^{ij}}I^j-\left(\sum^M_{j=1}{g^{ji}}\right)I^i 
\end{equation} 

\begin{equation} \label{GrindEQ__17_} 
\frac{dR^i}{dt}= \gamma \ I^i+\ \sum^M_{j=1}{h^{ij}}R^j-\left(\sum^M_{j=1}{h^{ji}}\right)R^i 
\end{equation} 

together with Eqs.\eqref{GrindEQ__9_}, \eqref{GrindEQ__11_} and \eqref{GrindEQ__13_}. The coefficients $f^{ij}$ etc. are not known a priori. They can be determined only after fitting (by trial and error method) the solutions for $S^i$, $I^i$ and $R^i$ to data. This will be an extremely time consuming process, fraught with uncertainties. Instead of proceeding along this line, we adopt a {\it stochastic plus deterministic} approach to deal with Eqs. \eqref{GrindEQ__15_} -- \eqref{GrindEQ__17_}. First we consider random exchanges of population (which could be $S$, $I$, $R$ or both) between the cities. We update the population after $\frac{M}{2}$ moves. Then, we solve Eqs. \eqref{GrindEQ__15_} -- \eqref{GrindEQ__17_} without the flow terms using Runge-Kutta-4(RK-4) method\cite{nr}. We consider three protocols of community transmission due to transportation of the population from one city to another:

\begin{itemize}
   \item {\bf CASE-I}: We consider $\frac{M}{2}$ numbers of pairwise exchanges of $I$ between the cities $i$ and $j$ by updating $I^{i}$ and $I^{j}$ to $\tilde{I}^{i}(={\eta }_I I^i+\ f_{I}\left(1-{\eta }_I\right)(I^i+I^j))$ and $\tilde{I}^{j}(={\eta }_I I^j+( 1-f_{I})\left(1-{\eta }_I\right)(I^i+I^j))$ respectively before we integrate the evolution equations as in earlier work\cite{ih1}. The exchange conserve the local populations, $\tilde{I}^i+\tilde{I}^j=I^i+I^j$ and $\tilde{N}^i+\tilde{N}^j=N^i+N^j$.    
   
   \item {\bf CASE-II}: We consider $\frac{M}{2}$ numbers of pairwise exchanges of $S$ (given by (a)) and $R$ (given by (b)) stochastically, with equal probabilities(=0.5) before we integrate the evolution equations. 
   
   (a) The susceptible populations in city $i$ (=$S^{i}$) and $j$ (=$S^{j}$) are updated to $ \widetilde{S^i}(={\eta }_S S^i+\ f_{S}\left(1-{\eta }_S\right)(S^i+S^j))$ and  $\widetilde{S^j}(={\eta }_S S^j+( 1-f_{S})\left(1-{\eta }_I\right)(S^i+S^j))$, conserving  $\tilde{S}^i+\tilde{S}^j=S^i+S^j$ and $\tilde{N}^i+\tilde{N}^j=N^i+N^j$. 
   
   (b) Similarly, the recovered populations in another set of randomly chosen cities $i$ (=$R^{i}$) and $j$ (=$R^{j}$) are updated to $ \widetilde{R^i}(={\eta }_R R^i+\ f_{R}\left(1-{\eta }_R\right)(R^i+R^j))$ and  $\widetilde{R^j}(={\eta }_R R^j+( 1-f_{R})\left(1-{\eta }_I\right)(R^i+R^j))$ conserving $\tilde{R}^i+\tilde{R}^j=R^i+R^j$ and $\tilde{N}^i+\tilde{N}^j=N^i+N^j$.
   
   \item {\bf CASE-III}: We consider $\frac{M}{2}$ numbers of pairwise inter-city exchanges of $S$, $I$ and $R$, stochastically, with equal probabilities($=\frac{1}{3}$),  before we integrate the evolution equations in each time step.
   
   (a) The susceptible populations in city $i$ (=$S^{i}$) and $j$ (=$S^{j}$) are updated to $ \widetilde{S^i}(={\eta}_S S^i+\ f_{S}\left(1-{\eta }_S\right)(S^i+S^j))$ and  $\widetilde{S^j}(={\eta }_S S^j+( 1-f_{S})\left(1-{\eta }_I\right)(S^i+S^j))$, conserving  $\tilde{S}^i+\tilde{S}^j=S^i+S^j$ and $\tilde{N}^i+\tilde{N}^j=N^i+N^j$. 
   
   (b) The infected population between the cities $i$ and $j$ are updated from $I^{i}$ and $I^{j}$ to $\tilde{I}^{i}(={\eta }_I I^i+\ f_{I}\left(1-{\eta }_I\right)(I^i+I^j))$ and $\tilde{I}^{j}(={\eta }_I I^j+( 1-f_{I})\left(1-{\eta }_I\right)(I^i+I^j))$ respectively before we integrate the evolution equations from $t$ to $t+1$ as in earlier work\cite{ih1}. The exchange conserve the local populations, $\tilde{I}^i+\tilde{I}^j=I^i+I^j$ and $\tilde{N}^i+\tilde{N}^j=N^i+N^j$. 
   
   (c) Similarly, the recovered populations in another set of randomly chosen cities $i$ (=$R^{i}$) and $j$ (=$R^{j}$) are updated to $ \widetilde{R^i}(={\eta }_R R^i+\ f_{R}\left(1-{\eta }_R\right)(R^i+R^j))$ and  $\widetilde{R^j}(={\eta }_R R^j+( 1-f_{R})\left(1-{\eta }_I\right)(R^i+R^j))$ conserving $\tilde{R}^i+\tilde{R}^j=R^i+R^j$ and $\tilde{N}^i+\tilde{N}^j=N^i+N^j$.
  
\end{itemize}

We use the connectivity parameter. $\eta=\eta_{S}=\eta_{I}=\eta_{R}\in [0,1]$ for simplicity, in the same way as in \cite{ih1}. The random fractions, $f_{S}$, $f_{I}$ and $f_{R}$ are random numbers $\in [0,1]$\footnote{2}. We test the initial condition as $N^{i}=1$, $R^{i}=0$, $I^{i}=10^-{4}$ and $S^{i}=N^{i}-I^{i}-R^{i}$ at $t=0$ for all $i$. Hence, the numbers that we quote in the paper is in the units of $N^{i}(0)(=1)$ as in \cite{ih1}.

\begin{figure}[h]
\includegraphics[scale=0.05]{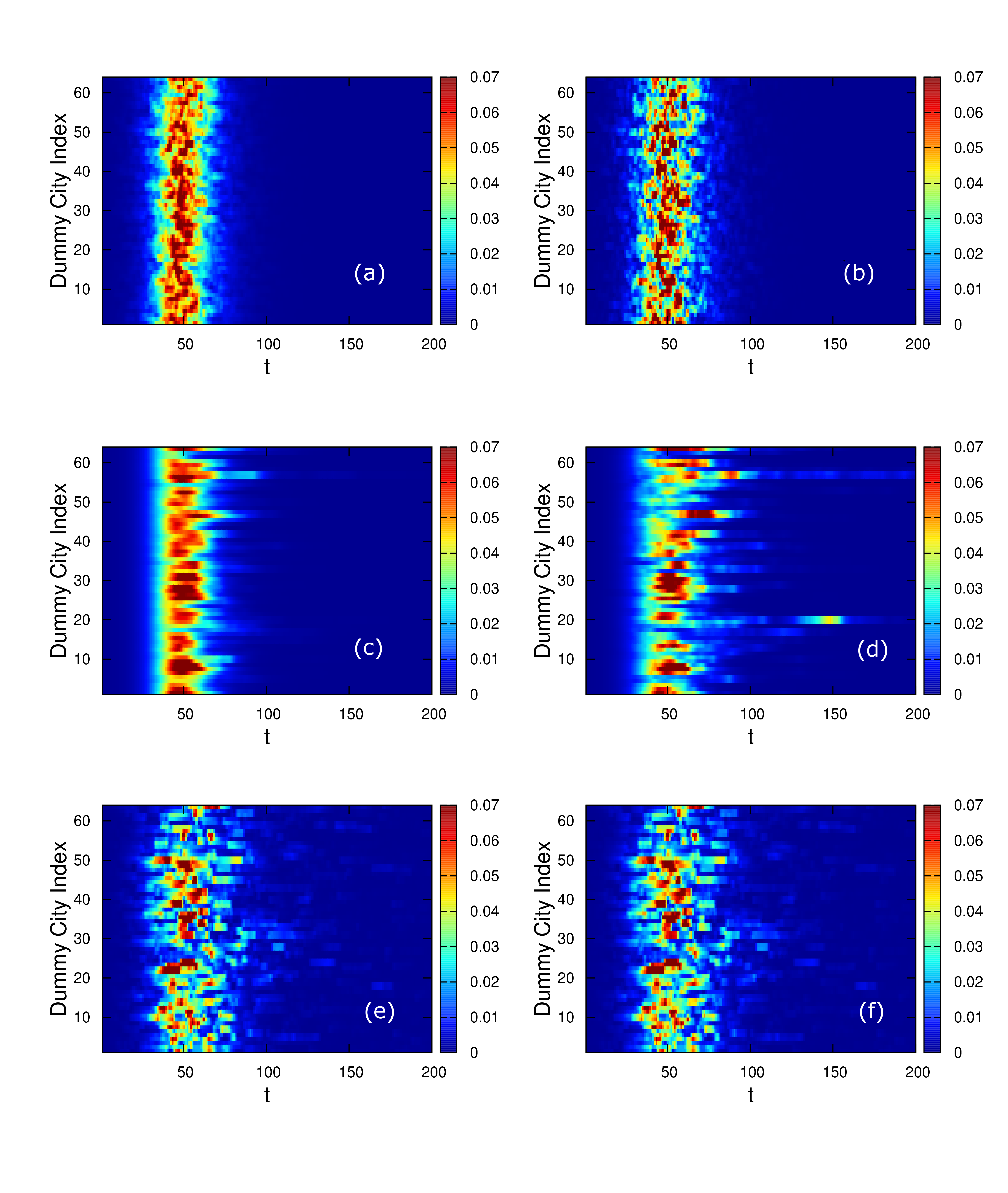}
\caption{Temporal evolution of Infected population (I) for different cities in a state consisting of $M=64$ cities (a-b) when only exchange of I is allowed with (a)$\eta=0.5$ (b)$\eta=0$, (c-d) when exchanges of S,R are allowed with equal probabilities with (c)$\eta=0.5$ (d)$\eta=0$ (e-f) when all the exchanges of S, I and R are allowed with equal probabilities with (e)$\eta=0.5$ (f)$\eta=0$.}
\end{figure}

In Fig.1, we show the evolution of the infected population, $I$ in the state consisting of $M=64$ cities for different cases. For CASE-I, the temporal evolution of $I(t)$ for all the cities are shown in Figs. 1(a) and (b) for $\eta=0.5$ and $\eta=0$ respectively when only exchange of infected population is allowed. In each of the cases, the peaks in infected population in different cities take place at different times as indicated by the diffusion of data points. The diffused line around $t\approx 50$ show the heterogeneity in the infection propagation. For CASE-II, the behavior changes for both the values of $\eta (=0.5$ and $0)$. We show this in Figs. 1(c-d) when the exchange of Susceptible (S) and Recovered (R) populations destabilize the infection propagation indirectly. This behaviour is quite surprising as the direct exchange of infected populations between the cities is ruled out. Yet, for $\eta=0$, the prominent secondary peaks in the infection appear at later times in more than one city suggesting strong heterogeneity due to increasing disorder which is may be due to the complex coupling that exists between $I$ and $S,R$. This becomes more prominent for CASE-III when all three types of populations commute between the cities (Fig. 1(c-d)). The different behaviour in the individual cities becomes more diffused for both $\eta=0.5$ (Fig. 1(e)) and $\eta=0$ (Fig. 1(f)). However, in this case, many peaks have significantly smaller numbers with respect to the other cases, suggesting the role of the exchange in mediating disorder.

\begin{figure}[h]
\includegraphics[scale=0.05]{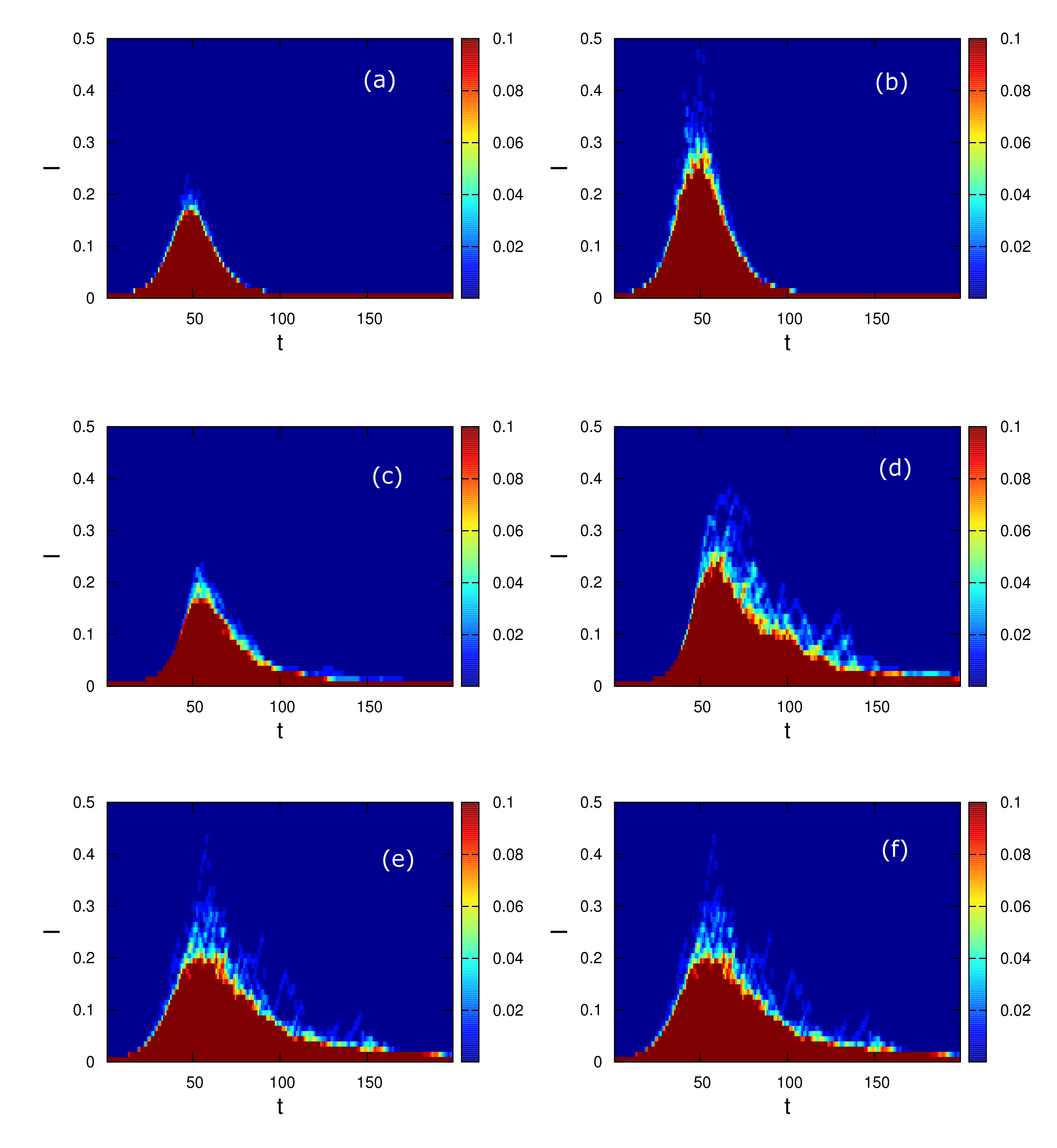}
\caption{Time evolution of the Probability distribution of Infected population (I) with $I$ at different $t$ in a state: (a-b) CASE-I with (a)$\eta=0.5$ (b)$\eta=0$, (c-d) CASE-II with (c)$\eta=0.5$ (d)$\eta=0$ (e-f) CASE-III with (e)$\eta=0.5$ (f)$\eta=0$.}
\end{figure}

We show the evolution of the Probability distribution functions of the infected population, $P(I,t)$ in Fig. 2 for the respective cases shown in Fig.1. For all the cases, $P(I;t)$ is a delta function at $t=0$, peaked at the initial values of the infection count at $t=0$, which is uniform and identical in the given scenario. For all the cases, we see that there is a spread in the distribution with increasing elapsed time upto a maximum limit. The maximal spread in the distribution behaves non-monotonically with increasing $t$, peaked at $t\approx 50$ for CASE-I with $\eta=0.5$ [Fig. 2(a)], when clear spread up to $I\approx 0.18$ is seen. This shoots up to $I\approx 0.3$ for the case with $\eta=0$ in Fig. 1(b). This suggests that the probability to have large peak infection increases with increasing $\eta$. This particular trend is also observed for CASE-II for both $\eta=0.5$ [Fig.2(c)] and $\eta=0$ [Fig.2(d)]. The large asymmetry with respect to the peak infection in the evolution process suggest that even the exchange of both $S$ and $R$ between the cities can impact the recovery process when the recovery process dominates at large $t$. At significantly large $t(= 125)$, there exist large probability to have very largely infected cities in the state which was not there earlier for CASE-I. For CASE-III, we see similar evolution for both $\eta=0.5$ and $\eta=0$. We also observe the similar asymmetric nature of the evolution that exist for CASE-II, emphasizing the slow recovery process.

\begin{figure}[h]
\begin{center}
\includegraphics[scale=0.12]{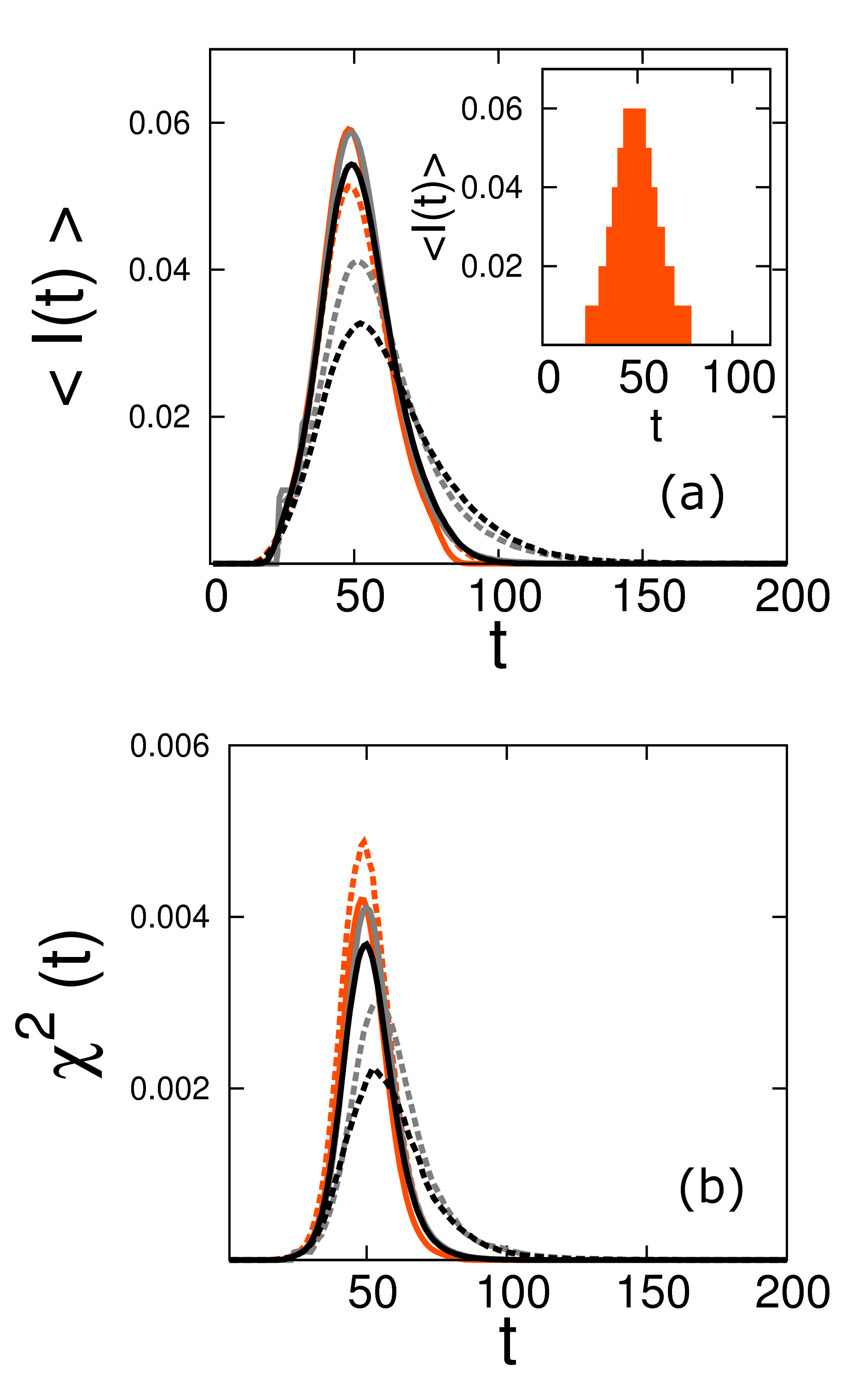}
\caption{(a) Evolution of the mean infected population $<I>$ with $t$ for CASE-I (orange), CASE-II (grey) and CASE-III (black) with $\eta=0.5$ (solid line), $\eta=0$ (dashed line). Inset. $<I>$ with $t$ for $\eta=1$. (b) $\chi^{2}(t)$ vs $t$ for the same cases as in (a).}
\end{center}
\end{figure}

We now discuss the evolution of the first ($<I>$) and second moment ($\chi^{2}$) of the probability distribution, $P(I;t)$, of infected populations in different cities, in Fig. 3. The instantaneous mean infected population, $<I(t)>(=\int dI IP(I;t))$ and the fluctuation contained in the distribution, $\chi^{2}(t)(=<I(t)^{2}>-<I(t)>^{2})$ for different cases are shown in Figs. 3(a) and 3(b) respectively. Inset. Fig: 3(a) shows the evolution of $<I>$ for cases with $\eta=1$ which are eventually identical as all the cities retain all three categories of population. 

For all these cases, $<I>$ has a peak at $t=50$ with value $<I>\approx 0.06$. As the exchanges take place, the nature of the evolution of $<I>$ changes. For CASE-I with $\eta=0.5$, the peak in $<I>$ has a similar value ($\approx 0.06$) with no significant asymmetry around the peak. For $\eta=0$, the peak in $<I>$ decreases by a small amount. Also small asymmetry around the peak for $<I>$ is observed. For CASE-II, we see a small pre-peak in $<I>$ at $t\approx 30$, before $<I>$ reaches the peak at $t\approx 50$ for $\eta=0.5$. For $\eta=0$, the height of the peak in $<I>$ decreases significantly (with value $\approx 0.04$). The distribution also shifts towards higher $t$ and develops large asymmetry with respect to the peak. For CASE-III, $<I>$ has a peak around $t\approx 50$ with notable asymmetry around $\eta=0.5$. Again, the height of the peak in $<I>$ decreases significantly for $\eta=0$. $<I>$ also shows very large asymmetry with respect to the peak. 

We now show the evolution of $\chi^{2}(t)$ for the respective cases of Fig.2. For CASE-I, $\chi^{2}$ has a peak at $t\approx 50$ with value ($\approx 0.004$) for $\eta=0.5$ and no significant asymmetry is observed. For $\eta=0$, the peak in $\chi^{2}$ grows higher. For CASE-II, the height of the peak in $\chi^{2}$ for $\eta=0.5$ has similar values for CASE-I. However, for $\eta=0$, the peak in $\chi^{2}$ decreases significantly and develops significant asymmetry for CASE-II. The end of epidemic is defined for $\chi^{2}\rightarrow 0$ when all the cities behave identically. This happens at a fairly large $t$ after the peak that $\chi^{2}$ develops. For CASE-III, peak in $\chi^{2}$ has values smaller than that for $\eta=0.5$ for CASE-II. For $\eta=0$, the height in $\chi^{2}$ decrease significantly and later slowly decays with $t$ and $\chi^{2} \rightarrow 0$ at significantly large $t$. 

\begin{figure}[h]
\includegraphics[scale=0.08]{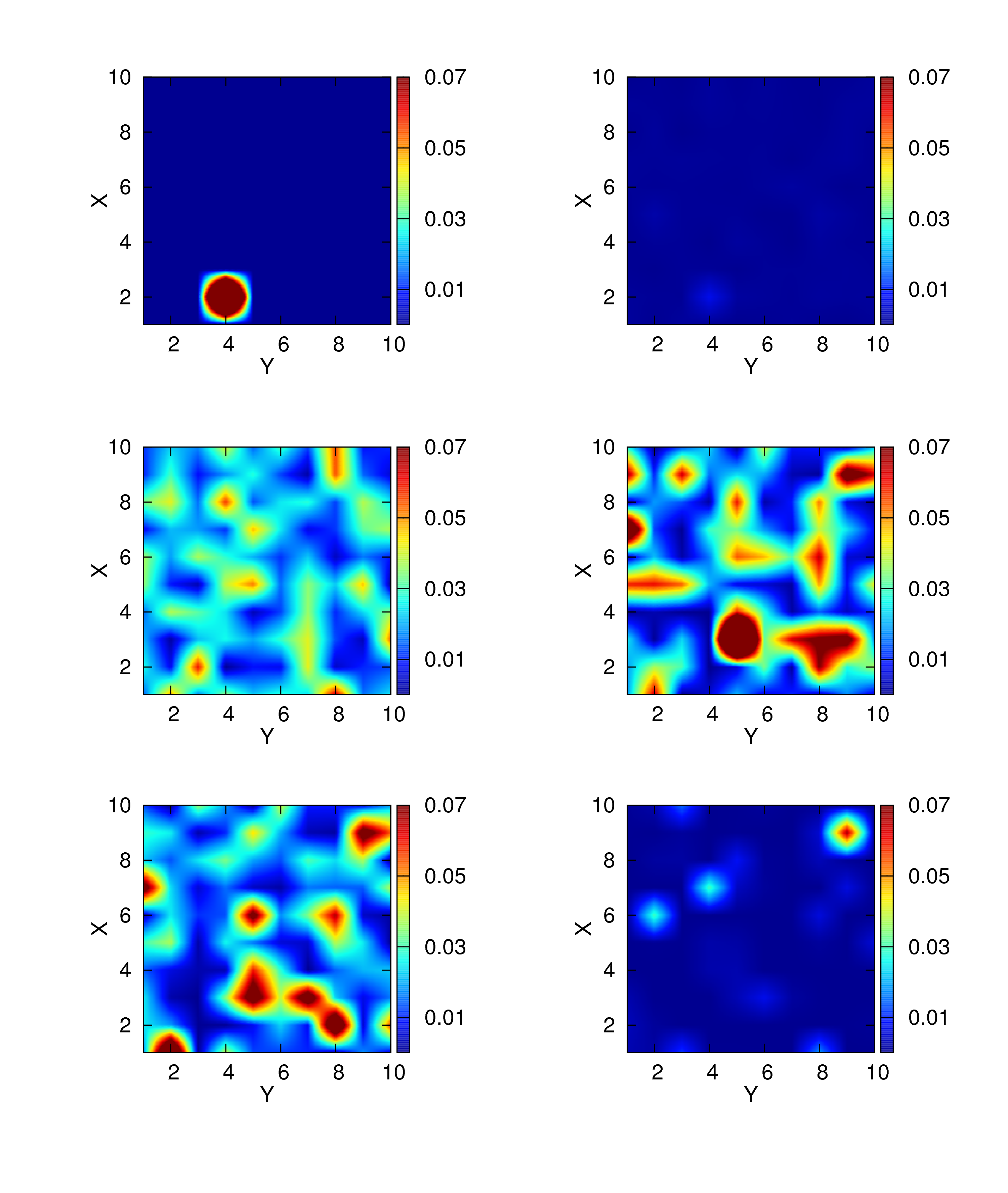}
\caption{Time evolution of the spatial maps of the infected populations in cities with nearest neighbour exchange at (a) $t=0$ (CASE-I) (b)$t=30$ (CASE-I)(c-e) $t=80$ (c) CASE-I (d) CASE-II (e) CASE-III (f)$t=120$ (CASE-III) with $\eta=0$.}
\end{figure}

We now test the three cases (CASE-I, CASE-II and CASE-III) in a state, consisting of $10X10$ cities ($X,Y\in L$), spatially located in a grid of length $L=10$ where each node is a representative of a city that follows similar equations of evolution. In contrast to earlier situations, the cities are now only allowed to exchange populations with the randomly chosen nearest neighbours following the identical governing rules (CASES-I,II and III) with $\eta=0$. In order to understand the dynamics of a realistic scenario, we choose a specific initial condition where only one randomly selected city has very large infected population ($I=0.2$) while others have very small infected cases (with $I=10^{-6}$) at $t=0$. 

The initial map of the infected populations in the cities is shown in Fig. 4 (a) which is identical for the three different trajectories following CASE-I, CASE-II and CASE-III. For all the three trajectories, the spatial maps are mostly identical for small $t$. We show the map of $I(X,Y)$ for $t=30$ in Fig. 4(b). The very high population at the initial city undergoes diffusion within this time window. No signature of the initial high value remains beyond $t=30$. The situation is mostly similar for all the trajectories for different CASES. We now show the maps of $I(X,Y)$ in Figs. 4(c-e) at $t=80$ for CASE-I, II and III respectively. $I(X,Y)$ contains many populated cities for CASE-I [Fig. 4(c)] while largely heterogeneous patches are seen in Fig. 4(d) for CASE-II. The map in Fig. 4(e) for CASE-III, has also many large patches of largely infected populations. At long times, $t=120$, there still exist a few relatively small patches of infected populations that decay very slowly for CASE-III [Fig. 4(f)] which is not present for CASE-I. For CASE-II, the situation is somewhat intermediate. 

\begin{figure}[h]
\includegraphics[scale=0.15]{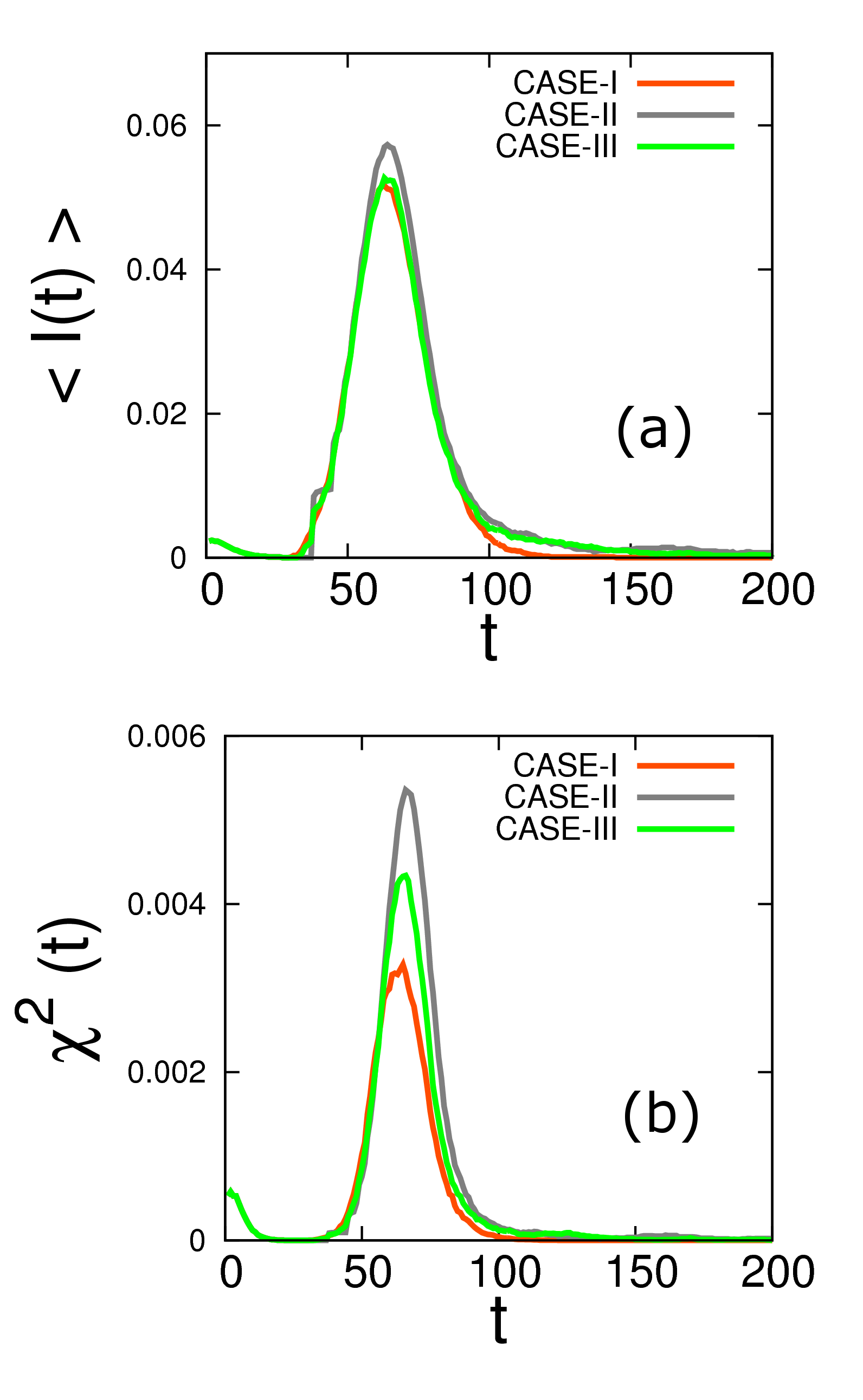}
\caption{(Colour Online) Time evolution of $<I>$ with $t$ for CASE-I (Orange Solid line), CASE-II (Grey solid line) and CASE-III (Green solid line) in a system that has one epicentre at $t=0$ in a two dimensional (10X10) spatial grid of cities with nearest neighbour interaction.}
\end{figure}

In Fig. 5. we now show the behavior of $<I>$ [Fig. 5(a)] and $\chi^{2}$ [Fig.5(b)] with $t$ for the three cases shown in Fig. 4. $<I>$ has a value $\neq 0$ at $t=0$ which is obvious. The initial decay in $<I>$ also persist up to $t\approx 30$. The peak in $<I>$ grows in a similar manner. However, for CASE-II, the height of the peak in $<I>$ is observed to be maximum while CASE-III has the slowest decay with increasing $t$[Fig. 5(a)]. The trend of data for $\chi^{2}$ is qualitatively similar to that for $<I>$[Fig. 5(b)]. The height of the maximum in $\chi^{2}$ is found to be maximum for CASE-II, at $t\approx 80$ due to existence of strong heterogeneity as seen in Fig. 4(d). For CASE-III, the decay in $\chi^{2}$ is observed to be the slowest which is also affirmed by the existence of the patches at $t=120$ in Fig. 4(f).  

In summary, we analyse the dynamics of interacting epicentres with intervention in a state using different protocols. The interaction was modelled using imposed disorder by exchanges of population of different kinds between the cities while the intervention parameter takes care of the fraction of population that a city retains before an exchange. We show that it is possible to control the heterogeneity in the infection propagation using the interplay between the intervention parameter and the complexity in the protocol. The control of heterogeneity essentially governs the fate of the containment strategy for the epidemic. The exchange model that we envisage here can easily be integrated with the existing models that takes care of many intermediate states between infection and recovery\cite{ind1,ger1,isical,ind2,lancet2,ind3}. Thus, the model is pivotal in shaping state-of-the-art non-pharmaceutical solutions to control the spread of the pandemic COVID-19.

The author is indebted to D. Syam for constant encouragements, numerous insightful discussions and comments on the draft. He is also acknowledged for a critical reading of the draft.


\begin{thebibliography}{}
\bibitem{footnote1}
In each timestep, we generate $M$ trial moves of pair-wise population exchanges. We only accept the moves if the distance between the randomly chosen cities is $=1$, without the periodic boundary condition. Also, we ensure that movement of a single category (also chosen stochastically) of population is allowed between the randomly chosen pair of cities, in a single move. For CASE-II, exchange of categories $S$ and $R$ is taken with equal probabilities (=0.5). The probability for each of the categories( $S$, $I$ and $R$), for CASE-III is $\frac{1}{3}$.
\bibitem{footnote2}
$f_{S}$, $f_{I}$ and $f_{R}$ are generated using independent uniform deviates ($\in [0,1]$).
\end{thebibliography}
\end{document}